\documentclass[preprint,prx,longbibliography]{revtex4-1}
\usepackage{graphicx}
\usepackage{comment}
\usepackage{amsmath,amssymb,amsthm}
\usepackage{bm}
\usepackage{verbatim}
\usepackage{float}

\begin{document}

\title{ Discrete dynamics versus analytic dynamics}

\author{S\o ren Toxvaerd}
\affiliation{DNRF centre  ``Glass and Time,'' IMFUFA, Department
 of Sciences, Roskilde University, Postbox 260, DK-4000 Roskilde, Denmark}
\date{\today}
\begin{abstract}
For discrete classical Molecular dynamics (MD) obtained  by the "Verlet" algorithm (VA) with the time increment $h$
there exists  a shadow Hamiltonian $\tilde{H}$ with energy $\tilde{E}(h)$,
for which the discrete particle positions lie on the analytic trajectories for $\tilde{H}$. Here we proof that there,
independent of such an analytic analogy, exists an exact hidden energy invariance $E^*$ for VA dynamics.
 The fact that the discrete VA dynamics
 has the same invariances as Newtonian dynamics  raises the question, which of the formulations that are  correct,
or alternatively, the most appropriate formulation of classical dynamics. In this context  the relation between
the discrete VA dynamics and the (general) discrete dynamics investigated by T. D. Lee [Phys. Lett.  $\textbf{122B}$, 217 (1983)]
is presented and discussed.
\end{abstract}
\maketitle

\section{Introduction}
 Molecular Dynamics (MD) generates the time evolution of $N$ classical mechanical particles by  discrete
time propagation. Almost all the MD are obtained by the "Verlet" algorithm (VA) \cite{Verlet} where
 a new position $\textbf{r}_i(t+\delta t)$ of the $i$'th particle with mass $m_i$
 at time $t+\delta t$ is obtained from the  force $\textbf{f}_i(t)$ and the two last
discrete positions
\begin{equation}
\textbf{r}_i(t+\delta t) = 2 \textbf{r}_i(t)- \textbf{r}_i(t-\delta t) + \frac{\delta t^2}{m_i}\textbf{f}_i(t).
\end{equation}
The algorithm is the central-difference expression for the mass times the acceleration of the particle which equals the
force $\textbf{f}_i$,
 and it appears in the literature with different names
(Verlet, leap-frog, velocity Verlet,..) \cite{tox1}. The algorithm is time reversible
 and symplectic, and the different reformulations
of the algorithm do not change  the  discrete time evolution and the physics obtained by the VA dynamics.

Mathematical investigations \cite{Sanz-Serna,Hairer,Reich} have proved
 the existence of a shadow Hamiltonian  $\tilde{H}$ \cite{tox2} for symplectic algorithms. The proof is   obtained by an asymptotic expansion,
 but  the series for the shadow
Hamiltonian does not converge in the general case. For a review of the asymptotic expansion,
its convergence and optimal truncation see \cite{Hairer1}.
 Only  the harmonic approximation, $E(1)$  of the first term in this expansion  is known explicitly \cite{tox1,tox2}.
But  inclusion of  $E(1)$ in the traditional obtained
zero order energy for    MD systems  with Lennard-Jones (LJ) particles
reduces the fluctuation  in   the energy by a factor of hundred for traditional values of $\delta t$ \cite{tox1}
and
 makes it possible to obtain the shadow energy,  $\tilde{E}$  of the   analytic dynamics with high precision.

 The VA algorithm  deviates, however, from all
other algorithms for classical dynamics by that $\textit{ momenta are not
 dynamical variables}$. Furthermore, the discrete VA dynamics for a harmonic
oscillator (DDHO), which  can be solved exactly, reveals that the DDHO not only has an
asymptotic expansion with an underlying analytic shadow Hamiltonian. But
the DDHO dynamics also has a $\textit{(hidden) energy invariance}, E^* $ which,
 independent of an existence of an analytic shadow Hamiltonian,
 is conserved step by step during the discrete time evolution. Below we show that this hidden invariance
 is a general quality of the discrete  VA dynamics,
 independent of the existence of a
shadow Hamiltonian, and that the discrete VA dynamics  has the same qualities and conserved
invariances as  analytic Newtonian dynamics.   In order to prove the existence of a hidden energy invariance we
must not make use of any analytic tools. This might seem to be a hopeless agenda, but on the other
hand the exact solution for a   discrete harmonic oscillator \cite{tox2}  makes no use of analyticity and
the exact solution  has an energy invariance $E^*$  which, in the analytic limit is equal to the energy of analytic Newtonian dynamics.

\section{The hidden energy invariance}

The kinetic energy in analytic  dynamics is obtained from the momenta.   
 The positions in Eq. (1) are the only dynamic variables in the discrete VA dynamics, i.e., the momenta, $\textbf{p}_i$,  are
 not. 
 Consequently, an expression  for the total momentum of the system requires a choice of an  expression for the
momentum $\textbf{p}_i$ of the $i'$th particle in terms of its positions.
The   sentences  "momenta",  "energy", "potential energy", "kinetic energy",
and "work" should be given by quotations in discrete VA dynamics to underline the fact, that the $N$ objects in the discrete dynamics only exercise mutual "irritations",
 or forces  $\textbf{f}_i(t_n)$ at their  positions at the discrete time $t_n=n\delta t$.
With the definition of the momenta
\begin{equation}
 \textbf{p}_i(t_{n-1},t_n) \equiv m_i \frac{\textbf{r}_i(t_n)-\textbf{r}_i(t_{n-1})}{\delta t}
\end{equation} it follows
immediately from the algorithm that the total momentum and angular momentum are conserved
for conservative systems with  $\sum_i^N \textbf{f}_i(t)=0$ \cite{tox4}. But the momenta and thereby the "kinetic energies" appear
$\textit{asynchronous}$ with the discrete positions and they are not a function of a single set of the discrete positions.

 The proof of an  invariance, equivalent
to the conserved energy in the analytic dynamics is more difficult, but it can be obtained
by  proving that there exists
 a hidden "energy" invariance, $E^*$,
 of the $N$ objects' dynamics with the change $\delta E^*_n=0$ by the discrete step which brings the $N$ positions  $\textbf{R}_n$
with the forces  $\textbf{F}_n$ at time $t_n$ to $\textbf{R}_{n+1}$ at
$t_{n+1}$.

For simplicity consider $N$ particles with equal masses $m_i=m$, and with the mass included in the
 discrete time increment, i.e. with $h \equiv \frac{\delta t}{\sqrt{m}}$.
A step with  discrete dynamics changes the "kinetic energy" of the system by $\delta K^*$ and its ability, $\delta U^*$, to perform a "work",
 $\delta U^*=-W^*$.
Since the momenta and thereby the kinetic energy is given by two sets of positions, a
 change in kinetic energy is given by three consecutive sets of positions.
The proof is obtained by consider two consecutive time steps. A new set of positions,  $\textbf{R}_{n+1}$ is obtained at the $n$'th time step
from the two previous sets, $\textbf{R}_{n-1}$, $\textbf{R}_n$ and the forces  $\textbf{F}_n$,
 by which  the change in the "kinetic energy" can be defined as 
\begin{equation}
\delta K_n^* \equiv  \frac{1}{2}  \left( \frac{\textbf{R}_{n+1}-\textbf{R}_{n}}{h} \right )^2-
 \frac{1}{2} \left( \frac{\textbf{R}_{n}-\textbf{R}_{n-1}}{h} \right )^2.
\end{equation}
The definition of the change in the kinetic energy for discrete VA dynamics 
 is consistent with the definition of the momenta, Eq. (2).

The forces $\textbf{F}_n$   bring the $N$ particles   to the positions $\textbf{R}_{n+1}$
 and with a change in the  ability, $\delta U^*_n$, 
  to perform a "work", $\delta U^*_n=-W_n^*$.  For the two steps we define the total change   
\begin{equation}
 2\delta U_n^*=-2 W_n^*  \equiv -\textbf{F}_n \cdot (\textbf{R}_{n+1}-\textbf{R}_{n-1}),
\end{equation}
and the discrete dynamics obeys the relation
\begin{equation}
\delta U_n^*+ \delta K_n^*=0.
\end{equation}

The proof starts by noticing that if one instead of the (NVE) dynamics,
 obtained by Eq. (1) with a constant time increment $h$,  
adjust the $(n+1)$'th time increment $h_n$ so $ W=0$,
 one obtains a geodesic step (NVU) \cite{nvu} to the positions $\textbf{R}_{n+1}(\textrm{U})$
which differs from  $\textbf{R}_{n+1}$. If
\begin{equation}
  -2 W(\textrm{U})_n = -\textbf{F}_n \cdot (\textbf{R}_{n+1}(\textrm{U})-\textbf{R}_{n-1})=0
\end{equation}
is inserted in the Verlet algorithm; Eq. (1) 
\begin{equation}
 \textbf{R}_{n+1}(\textrm{U})=2 \textbf{R}_{n}-\textbf{R}_{n-1}+ h_n^2 \textbf{F}_{n},
\end{equation}
  one
obtains  an expression for $h_n^2$ at the  NVU step at time $t_n$  \cite{nvu}
\begin{equation}
 h_n^2=- 2 \frac{\textbf{F}_n \cdot(\textbf{R}_{n}-\textbf{R}_{n-1})}{\textbf{F}_n^2}.
\end{equation}
 I.e. instead of propagating the system  the $n$'th step with the constant time increment
$h$, the increment $h_n$ is adjusted to ensure that the system ability to perform a work  is unchanged.

 The NVU step at time $t_n$ updates the position to $\textbf{R}_{n+1}(\textrm{U})$ and
 with the  geodesic invariance: the constant length of the steps \cite{nvu}
\begin{equation}
(\textbf{R}_{n+1}(\textrm{U})-\textbf{R}_{n})^2=(\textbf{R}_{n}-\textbf{R}_{n-1})^2,
\end{equation}
which is  obtained  by   rearranging and  squaring Eq. (7)
\begin{eqnarray}
(\textbf{R}_{n+1}(\textrm{U})-\textbf{R}_{n})^2 &=& \left( \textbf{R}_n-\textbf{R}_{n-1}
- 2 \frac{\textbf{F}_n \cdot(\textbf{R}_{n}-\textbf{R}_{n-1})}{\textbf{F}_n^2} \textbf{F}_n \right )^2\\
 & = &  (\textbf{R}_n-\textbf{R}_{n-1})^2,
\end{eqnarray}
i.e.  with the change in "kinetic energy"
\begin{equation}
  \delta K(\textrm{U})_n =  \frac{1}{2} \left( \frac{\textbf{R}_{n+1}(\textrm{U})- \textbf{R}_{n}}{h} \right )^2-
 \frac{1}{2} \left( \frac{\textbf{R}_{n}-\textbf{R}_{n-1}}{h} \right )^2=0.
\end{equation}
So the NVU step to $\textbf{R}_{n+1}(\textrm{U})$  obeys
\begin{equation}
   \delta U(\textrm{U})_n= \delta K(\textrm{U})_n=0. 
\end{equation}

 We are now able to proof the existence of an  "energy"
invariance (Eq. (5)) by the VA dynamics, Eq. (1). The proof can e.g. be obtained
 by deriving the difference between the NVU and the NVE step at $t_n$.
 The new positions $\textbf{R}_{n+1}$ and $\textbf{R}_{n+1}(\textrm{U})$ are both obtained
from $\textbf{R}_{n-1}, \textbf{R}_{n}$ and $\textbf{F}_{n}$, but with different time increments.
With NVE
\begin{eqnarray}
 2 \delta U_n^*= -2 W_n^* & = &-\textbf{F}_n \cdot (\textbf{R}_{n+1}-\textbf{R}_{n-1}) \nonumber \\
                           & = &- \textbf{F}_n \cdot (\textbf{R}_{n+1}-\textbf{R}_{n+1}
(\textrm{U})+\textbf{R}_{n+1}(\textrm{U})-\textbf{R}_{n-1}) \nonumber \\
                     & = & -\textbf{F}_n \cdot (\textbf{R}_{n+1}-\textbf{R}_{n+1}(\textrm{U})).
\end{eqnarray}

The difference $\textbf{R}_{n+1}-\textbf{R}_{n+1}(\textrm{U})$
 can be obtained from the Verlet algorithm, Eq. (1)  and the NVU algorithm, Eqn. (7) and (8),
and gives
\begin{equation}
 2 \delta U_n^*= -2 W_n^* = -(h^2 \textbf{F}_n^2+ 2 \textbf{F}_n \cdot(\textbf{R}_{n}-\textbf{R}_{n-1})). 
\end{equation}
The change in the "kinetic energy"  is obtained from the NVE algorithm Eq. (1)
\begin{eqnarray}
 2\delta K_n^* & = &  \left( \frac{\textbf{R}_{n+1}-\textbf{R}_{n}}{h} \right )^2 - 
 \left ( \frac{\textbf{R}_{n}-\textbf{R}_{n-1}}{h} \right )^2 \nonumber\\
 & = &  \frac{(\textbf{R}_{n}-\textbf{R}_{n-1})^2+ h^4 \textbf{F}_n^2+ 2 h^2 \textbf{F}_n \cdot(\textbf{R}_{n}-\textbf{R}_{n-1})}{h^2}
   -  \left ( \frac{\textbf{R}_{n}-\textbf{R}_{n-1}}{h} \right )^2  \nonumber \\
 & = & h^2 \textbf{F}_n^2+ 2 \textbf{F}_n \cdot(\textbf{R}_{n}-\textbf{R}_{n-1}), 
\end{eqnarray}
 and  Eq. (5) for the discrete dynamics is obtained from Eqn. (15) and (16).

The change in kinetic
energy in discrete dynamics  must necessarily be obtained from two consecutive steps
 and the change in the systems ability to perform
a work is consistently obtained from the same sets of positions. But,
 by  eliminating $\textbf{R}_{n-1}$ in  Eq. (4), one obtains an expression
 for the change in  the ability per time step, $\delta U_n^*$ 
\begin{eqnarray}
  \delta U_n^*= - \frac{1}{2}\textbf{F}_n \cdot (\textbf{R}_{n+1}-\textbf{R}_{n-1}) \nonumber\\
       = -\textbf{F}_n \cdot (\textbf{R}_{n+1}-\textbf{R}_{n})+ \frac{h^2}{2} \textbf{F}_n^2.
\end{eqnarray}

In the  Newtonian dynamics the existence of a potential energy  state function, $U(\textbf{R})$, is ensured by that
the total work done around any closed circuit from $\textbf{R}$  is zero \cite{Goldstein}. 
The VA dynamics is started from two  sets of positions $\textbf{R}_{0}$ and $\textbf{R}_{1}$, and the time increment $h$. 
The ability to perform a discrete work $U^*(\textbf{R})$ is also a state function
  and it plays the same role as the
potential energy $U(\textbf{R})$ for analytic Newtonian dynamics. Consider any discrete closed  sequence of positions  
generated with VA dynamics with the time increment $h$ and
 which starts and ends with  the same two  configuration $\textbf{R}_0$ and $\textbf{R}_{1}$.
 The total change in the
kinetic energy is
\begin{equation}
\sum_{i=1}^{i=n} \delta K^*_i= \frac{(\textbf{R}_2-\textbf{R}_1)^2}{h}-\frac{\textbf{(R}_1-\textbf{R}_0)^2}{h}+
\frac{(\textbf{R}_3-\textbf{R}_2)^2}{h}-\frac{\textbf{(R}_2-\textbf{R}_1)^2}{h}+...+\frac{\textbf{(R}_1-\textbf{R}_0)^2}{h}=0. 
\end{equation}
The start ability is $U^*(\textbf{R}_1)$, 
and since all the terms in
$\sum_{i=1}^{i=n} (\delta U^*_i+\delta K^*_i)$ are zero accordingly to Eq. (5)  it implies  that
\begin{equation}
\sum_{i=1}^{i=n} \delta U^*_i=0.
\end{equation}
   
The energy invariance 
\begin{equation}
E^*=U_n^*(\textbf{R}_n)+K_n^*(\textbf{R}_{n-1},\textbf{R}_n)
\end{equation}
 is given by the start condition
for the discrete dynamics and it differs from the energy invariance of Newtonian dynamics. It is a state function, and due to the
discrete dynamics it depends on two consecutive sets of the positions, $\textbf{R}_{n-1},\textbf{R}_n$,
 instead of the energy invariance in Newtonian dynamics which depends on
the positions $\textbf{R}(t_n)$ and the momenta $\textbf{P}(t_n)$ at the same time $t_n$. The two
invariances are, however, equal in the analytic limit \cite{Goldstein}
\begin{equation}
\lim_{h \rightarrow 0} \left ( \delta U_n^*+ \delta K_n^* \right ) =
 -\textbf{F}(t_n) \cdot \delta \textbf{R}(t_n)+ \delta K(t_n) +\mathcal{O}(h^2)  =0,
\end{equation}
where the  term $\mathcal{O}(h^2)=\frac{ h^2}{2} \textbf{F}(t_n)^2$ is the $\textit{total deviation}$ from the Newtonian dynamics.

  The  invariance, Eq. (20) does not depend on a convergence of an asymptotic expansion,
 and it differs also from  the  shadow energy for the  shadow Hamiltonian
by that, although the change contains two terms, the expressions for their changes do not make use of a potential, but only of the  forces
and the discrete positions.
It is obtained by noticing that, with a suitable definition of  the "work" and kinetic energy, 
 $\delta U^*_n=-\delta K^*_n$, and by formulating the requirement   that $U_n^*(\textbf{R}_n)$ is a state
function. The derivation is a  copy of the
derivation of the energy invariance for Newtonian dynamics \cite{Goldstein}. In Thermodynamics the First law of thermodynamics
is formulated exactly in the same manner, but as a basic assumption  of that the energy function is a state function consisting
of two terms  which change by work and kinetic energy exchanges, and the present formulations is the 
  corresponding formulation of
the energy conservation in dynamics  and thermodynamics for discrete VA dynamics.

\subsection{Energy conservation in MD with VA dynamics}

The formulation of energy in discrete VA dynamics by the ability $U^*(\textbf{R})$ to perform a discrete work instead of the potential
energy  $U(\textbf{R})$ works equally well as the traditional formulation.
Molecular Dynamics simulations with VA for $N$ particles are obtained from two consecutive start sets of positions, $\textbf{R}_{0}$ and
 $\textbf{R}_{1}$, and these positions define
not only  the total dynamics evolution, but also the mean value of
 traditional zero order energy, $<E(0)_n>$,  the accurate first order estimate of
 the shadow energy, $\tilde{E}_n \approx E(0)_n+h^2 E(1)_n$ of the
underlying analytic dynamics and the exact energy invariance $E^*$. Since  the change in $E^*$ is given in the same manner as the
 energy conservation by the First law of thermodynamics, we  need to define  a  "start ability", $U_1^*$ for the discrete dynamics.
But  the discrete VA dynamics  differs  in fact neither from the analytic counterpart at this point.
 In principle we could obtain the ability $U_1^*$ at the start of
the simulation
by determining  the discrete work performed by bringing the particles from infinite separations 
 via $\textbf{R}_{0}$ to the positions $\textbf{R}_{1}$.  In  the thermodynamics one defines, however, a standard
 state of  energy (enthalpy), and here we will use
the   potential energy $U_1(\textbf{R}_1)$ at the positions $\textbf{R}_1$ and the accurate estimate,  $\tilde{E}_1$ at the start of the dynamics, and obtain 
\begin{equation}
U_1^*=U_1(\textbf{R}_1),
\end{equation}  
and the energy invariance
\begin{equation}
 E^*(t_{n+1})=\tilde{E}_1+\sum_{i=1}^{i=n} \left [\delta U_n^* +
 \delta K_n^* \right ]
\end{equation}
with
\begin{equation}
 \delta U_n^*= -\textbf{F}_i \cdot(\textbf{R}_{i+1}-\textbf{R}_{i}) +\frac{h^2}{2}\textbf{F}_i^2. 
\end{equation}
The energy evolution by MD in double precision arithmetic with VA was determined for two systems. In the first a liquid system  of  $N=2000$ LJ particles
at the density $\rho=0.80 \sigma^3$ was calibrated at the temperature $kT/\epsilon=1$. The thermostat \cite{tox4} was switched off and
the energy evolution in the next ten thousand time steps with $h=0.005$ was obtained.
\begin{figure}[h!]
\begin{center}
\includegraphics[width=6.cm,angle=-90]{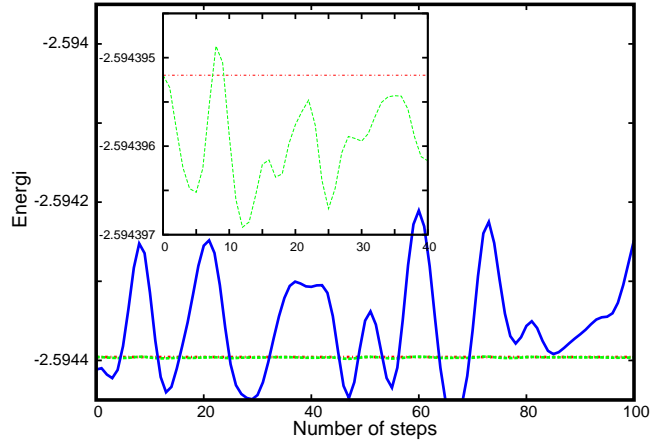}
\caption{ Discrete  energies of a LJ system with the Verlet algorithm at $T,\rho$ = 1.0, 0.80 and for
 $h$=0.005.  Blue solid line: Traditional energy estimate $E(0)_n$;
 green dashed line: "Shadow" energy $\tilde{E}_n \approx E(0)_n+h^2 E(1)_n$;
 red dash-dotted line: the  energy invariance $E^*$. The inset shows  $\tilde{E}_n$ and  $E^*$.
}
\end{center}
\end{figure}
\begin{figure}[h!]
\begin{center}
\includegraphics[width=6.cm,angle=-90]{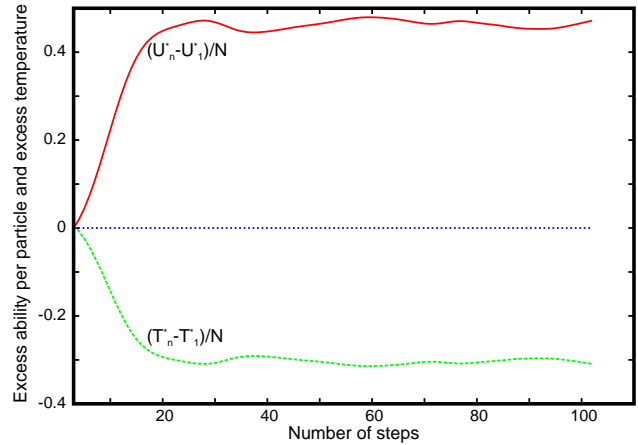}
\caption{ Excess discrete  energies end temperatures  of a  non-equilibrium system of $N=2048$ LJ particles  at spontaneous melting .
  Red line: the excess ability per particle $(U_n^*-U_1^*)/N$; green  long dashed: excess temperature  $T^*_n-T^*_1$;
blue short dashes: the constant energy invariance $E^*_n=E^*_1$.
}
\end{center}
\end{figure}
  Figure 1 shows the
energy evolution $E(0)_n, \tilde{E}_n$ and $E^*_n$ for the first hundred time steps.
 The first order estimate of $\tilde{E}_n$ (green dashes) improves the accuracy of the energy determination
with a factor of hundred, the energy invariance  $E^*$ is exact (see inset).

The constant VA dynamics is obtained from two sets of positions, $\textbf{R}_{0},\textbf{R}_{1}$ and $h$ and these start values can not contain information
about whether the system is in equilibrium or not. 
In order to obtain the evolution of the kinetic energy $K_n^*$ and the ability $U_n^*$ in a non-equilibrium system
with VA dynamics a system was started with two sets of positions
which correspond to a non-equilibrium state. The  non-equilibrium state was obtained for a system of  $N=2048$ LJ particles in a fcc solid
 at 
   $kT/\epsilon=1$  and  density $\rho=1.009 \sigma^3$ ( density
of coexisting solid at $kT/\epsilon=1$ \cite{sol}) by spontaneously expanding the positions to the
 density $\rho=0.80 \sigma^3$  by a scaling of all the positions  $\textbf{R}_{0},\textbf{R}_{1}$. A LJ systems equilibrium state
at the density  $\rho=0.80 \sigma^3$ is a liquid.
The fcc ordered system melted spontaneously , and the conservative systems temperature  decreased
 according to the Second law of Thermodynamics. The  change in the  temperature  at the spontaneous melting is shown with green dashes
in Figure 2. The temperature decreased from  $T^*_1=1.$ within 20-40 time steps to $T^* \approx 0.7$. The differences between $T^*_n$
and the temperature  $\tilde{T}_n$,  obtained  for the shadow Hamiltonian \cite{tox4} are of the order $10^{-7}$, and they
are not visible on the figure. The  decrease in the spontaneous temperature
 at the  melting was balanced by a corresponding increase in the ability (red line),
and the energy invariance $E^*$ (blue small dashes) was constant in the conservative system.
The two MD simulations (Figure 1 and Figure 2) demonstrate that the traditional and the present (discrete) energy concept work equally well. 

\section{ Discrete dynamics versus analytic dynamics}

The discrete VA dynamics has the same invariances as Newtonian dynamics 
and it raises the question: Which of these formulations that are  correct,
or alternatively, the most appropriate formulation of classical dynamics? In this context  T. D. Lee
in  1983  wrote a paper \cite{Lee1} entitled, "Can Time Be a Discrete Dynamical Variable?";
which led to a series of publications by Lee and collaborators  on the formulation of fundamental
dynamics in terms of difference equations, but with exact invariance under
continuous groups of translational and rotational transformations.
Quoting Lee \cite{Lee2}, he "wish to explore an alternative point of view: that physics
should be formulated in terms of difference equations
and that these difference equations could exhibit all the desirable symmetry properties and conservation laws".
 Lee's analysis covers not only classical mechanics \cite{Lee1}, but also non relativistic quantum mechanics
 and relativistic quantum field theory  \cite{Lee3}, and  Gauge theory and Lattice Gravity \cite{Lee2}.
The discrete dynamics is obtained by treating positions and time, $\textit{but not momenta}$, as a discrete dynamical variables, and he
obtained a conserved  (mean) "energy"  over consecutive time intervals of different lengths.
But  according to Lee \cite{Lee1} in his formulation of discrete mechanics, "there
is a $\textit{fundamental length}$ or time $l$ (in natural units). Given any time interval
 $T=t_{\textrm{f}}-t_0$, the total number $N$ of discrete points that define the
trajectory is given by the integer nearest $T/l$."
 
 The analogy between Lee's formulation
of discrete dynamics  and  VA dynamics is striking. For the VA dynamics  one uses a  $\textit{unit time increment, h}$, and  the momenta
are not dynamical variables and they have no impact on the
 discrete dynamics \cite{tox1}.  The fundamental length and time  in  quantum electrodynamics are
the  Planck length $l_{\textrm{P}} \approx 1.6 \times 10^{-35}$m and
 Planck time $t_{\textrm{P}} \approx 5.4 \times 10^{-44}$ s \cite{Garay},
and they are immensely smaller than the length unit (given by the floating point precision)
 and time increment used in MD to generate the classical discrete dynamics.
 But the analogy implies that the  discrete VA dynamics obtained by MD is the "continuation" of the
Lee's discrete quantum dynamics for a fundamental length of time $t_{\textrm{P}}$,
 as is the analytic classical dynamics of the traditional quantum mechanics,
given by the Wigner expansion \cite{Wigner}.

The discrete non relativistic quantum mechanics  is  obtained by Lee using Feynman's path
 integration formalism, but for  discrete positions and a corresponding 
discrete action,
\begin{equation}
  \mathcal{A}_D = \sum_{n=1}^{N+1} \left [ \frac{(\textbf{R}_{n}-\textbf{R}_{n-1})^2}{2(t_n-t_{n-1})}+
(t_n-t_{n-1}) \overline{V(n)} \right ]
\end{equation} 
where $\textbf{R}_{N+1}$ is the end-positions  at time $t_{N+1}$ and the  minimum  of $ \mathcal{A}_D$ determines the classical path.
The  action is a sum over products of time increments and "kinetic energies" $K_n^*$, and Lee has used the symbol
 $ \overline{ V(n)}$, for the 
  average of "potential energy" in the
time intervals $[t_{n-1},t_n]$.
The momenta for all  the paths, given by the discrete nodes 
  $\textbf{R}_{1},..,\textbf{R}_{N+1}$ are  obtained from differences,
 $\textbf{R}_{n}-\textbf{R}_{n-1}$, so the classical VA discrete trajectory is the
 classical limit path for  discrete quantum mechanics
 with $h=t_{\textrm{P}}$, as  the classical Newtonian trajectory is for the traditional quantum mechanics.
There is, however, one important difference between the analytic and the discrete dynamics.
 The momenta  $\textit{for all the paths}$ in the discrete quantum dynamics are obtained by a difference between
discrete sets of positions and they are all $\textit{asynchronous}$ with the positions. So
the Heisenberg uncertainty is a trivial consequence of  a discrete quantum electrodynamics with
a fundamental length of time $t_{\textrm{P}}$.

Lee motivates his reformulation of the analytic dynamics in  the Introduction in \cite{Lee3} by the difficulties of formulating
a general $\textit{unifying}$
theoretical model for dynamics and with the Concluding remarks in \cite{Lee2} that (he tries to explore the opposite viewpoint):
"Difference equations are more fundamental, and differential equations are regarded as approximations".
 The difference in the energy between the
 analytic energy and the energy obtained by Eq. (21) for discrete electrodynamics with a unit time increment $t_P$ is of the
order $t_P^2$, and it is  absolute marginal. The Heisenberg uncertainty between positions and momenta is of the order $t_P$  and
this uncertainty  is
an inherent  quality of  discrete dynamics with a  fundamental length of time $t_{\textrm{P}}$.
The discrete classical VA dynamics is fundamentally different from analytic Newtonian dynamics, but has the same invariances and
the dynamics is obtained equally well by both methods.
But, on the other hand the traditional quantum mechanics  is in all manner  fully appropriate
and justifies no  revision of the formulation, and an eventual revision of the dynamics
 must be justified by other facts than conservation of the energy by classical Molecular Dynamics
simulation with the VA algorithm.

\acknowledgments
The author acknowledges useful discussions with Ole J Heilmann and Jeppe C Dyre.
The centre for viscous liquid dynamics ``Glass and Time'' is sponsored by the Danish National Research Foundation (DNRF) grant No. DNRF61.
$ $\\

\end{document}